\documentclass[%
 reprint,
superscriptaddress,
 amsmath,amssymb,D_1
 aps,
]{revtex4-1}

\usepackage{graphicx}
\usepackage{dcolumn}
\usepackage{bm}
\usepackage{mathrsfs}
\usepackage{amsmath}
\usepackage[usenames,dvipsnames]{color} 
\usepackage{amssymb}

\usepackage{graphicx}

\usepackage{subfigure}

\usepackage{hyperref}
\usepackage[mathlines]{lineno}

\begin{document}

\preprint{APS/123-QED}

\title{Non-Gaussian ground-state deformations near a black-hole singularity}

\author{Stefan Hofmann}
 \email{stefan.hofmann@physik.uni-muenchen.de}
  \affiliation{Arnold Sommerfeld Center for Theoretical Physics, Theresienstra{\ss}e 37, 80333 M\"unchen\\}

\author{Marc Schneider}
 \email{marc.schneider@physik.uni-muenchen.de}
  \affiliation{Arnold Sommerfeld Center for Theoretical Physics, Theresienstra{\ss}e 37, 80333 M\"unchen\\}



\date{\today}

\begin{abstract}
The singularity theorem by Hawking and Penrose qualifies Schwarzschild black-holes as geodesic incomplete
space-times. Albeit this is a mathematically rigorous statement, it requires an operational framework that 
allows to probe the space-like singularity via a measurement process. Any such framework necessarily 
has to be based on quantum theory. As a consequence, the notion of classical completeness needs to be 
adapted to situations where the only adequate description is in terms of quantum fields in dynamical 
space-times. It is shown that Schwarzschild black-holes turn out to be complete when probed by 
self-interacting quantum fields in the ground state and in excited states. The measure for populating 
quantum fields on hypersurfaces in the vicinity of the black-hole singularity goes to zero towards
the singularity. This statement is robust under non-Gaussian deformations of and excitations relative to the 
ground state. The physical relevance of different completeness concepts for black holes is discussed.

\begin{description}
\item[PACS numbers]
04.20.Dw, 04.62.+v, 04.70.-s, 11.10.-z
\end{description}
\end{abstract}

\pacs{04.20.Dw, 04.62.+v,04.70.-s,11.10.-z}
\keywords{Suggested keywords}
\maketitle


\section{\label{sec:level1}Introduction}
The singularity theorem \cite{hawk70} by Hawking and Penrose 
identifies Schwarzschild black-holes as incomplete in a precise 
sense: Black holes incorporate a space-like singularity where 
null and time-like geodesics end prematurely, referring to 
classical point particles that reach these endpoints in a finite time,
because their potential is bounded from above \cite{reed75}. 
This relates the geometric completeness concept to the usual notion 
of potential completeness. The latter can be lifted to quantum mechanical
completeness, which implies the existence of a unique evolution in
compliance with unitarity. Unitarity remains the relevant completeness
criterion in static space-times and extends to encompass the relativistic
domain of a single particle. 

In this context, Horowitz and Marolf \cite{hor95} were the first to point out
that geodesic (and hence potential) incompleteness does not necessarily imply
unitarity violation. They gave examples of static space-times with time-like
singularities that nevertheless qualified as complete from a quantum mechanical
perspective. This was conceptually promising since quantum field theory in a static,
globally hyperbolic space-time admits a consistent description of a single
relativistic particle, as was shown by Ashtekar and Magnon in \cite{ash75}. 
And it was practical, since Wald \cite{wald78} showed that 
the dynamics of a Klein-Gordon scalar field in arbitrary static space-times 
could be examined by asking whether the spatial part of the wave operator
admits a self-adjoint extension.     

Dynamical space-times in general, however, require a quantum theory with fields as local
bookkeeping devices. The interior of Schwarzschild black-holes is a dynamical space-time,
even though the exterior is thought of as being static.
A strictly unitary evolution is no longer necessary, even in the 
absence of interactions. As a consequence, a new criterion for quantum completeness is
required that reduces to the classical one (and its quantum mechanical descendant)
in appropriate limits, but extends to quantum field theory.
While unitarity reflects the symmetry underlying quantum mechanical evolution,
the logically more potent concept is state normalisation. Unitarity is replaced
by stability, which demands a valid probabilistic interpretation instead of a conserved norm.
So stability requires only a semigroup of contractions \footnote{This corresponds to relaxing 
the requirements posed on evolution generators to linear unbounded operators that are closed, 
have a dense domain, whose resolvent set contains 
all positive reals and whose resolvent operator, which can be thought of as the Laplace transform
of the semigroup, is bounded from above for each element of the resolvent set by its inverse.        
}.
At the intuitive level, stability ensures a probabilistic interpretation of the quantum
system in a background dynamical space-time. 
Such a stability notion clearly reflects on the completeness of the background space-time as scrutinised 
by quantum fields. 

Stability investigations are usually pursued in the asymptotic framework pertinent to scattering theory,
which is neither an option in generic space-times, nor is it practical given that instabilities are anchored
in regions near classical singularities. In this situation, the Schr\"odinger representation of quantum field
theory turns out to be extremely useful, since it conveniently allows to investigate stability at finite times. 
Based on this framework, the following completeness criterion \cite{sh15} has recently been suggested:
A globally hyperbolic space-time is called quantum complete to the left with respect to a free field theory, 
if its Schr\"odinger wave functional can be normalised at an initial time $t_0$,
and if the normalisation is bounded from above by its initial value for all $t\in(0,t_0)$.

In a previous article \cite{sh15} it has been shown that Schwarzschild black-holes are quantum complete.
Here, the notion of quantum completeness is extended to include non-Gaussian deformations
of the ground state induced by self-interactions and excited states. As will be shown, all 
generalisations respect the concept of quantum completeness as suggested above. 
Concrete calculations are presented for a Schwarzschild black-hole populated by real scalar fields with quartic 
self-interactions. The wave functionals for the ground state as well as for arbitrary excited states
are investigated near the black-hole singularity. 
The main result is that Schwarzschild black-holes are quantum complete even if self-interactions and
excitations are permitted.
The different completeness concepts employed in physics are logically consistent in their respective 
domains of validity. Their physical relevance for black-hole interiors is discussed
in detail.

\section{\label{sec:level2}Geometric Preliminaries}
Let us first clarify our conventions. Consider the space-time $(\mathcal{M},g)$ with
$\mathcal{M}:=\mathbb{R}\times\mathbb{R}^+\times S^2$,
where $S^2$ is the unit two-sphere. The projections $t: \mathcal{M}\rightarrow \mathbb{R}$
and $r: \mathcal{M}\rightarrow \mathbb{R}^+$ are called Schwarzschild time and Schwarzschild radius, respectively. 
The Schwarzschild function $h(r):=1-r_{\rm g}/r$ is increasing from minus infinity at $r=0$ to one as $r$
approaches plus infinity, passing through zero at $r=r_{\rm g}$. Here, $r_{\rm g}=2 M$ denotes the 
gravitational radius of a source of mass $M$.  
The physical conditions implied by a static and spherical symmetric source in vacuum, 
supplemented with asymptotic fall-off conditions,
give rise to two warped product space-times,
the Schwarzschild exterior space-time $\mathcal{E}:=P_>\times_r S^2$, with $P_>$ denoting the region
$r>r_{\rm g}$ in the $(t,r)$-half plane $\mathbb{R}\times\mathbb{R}^+$, and the Schwarzschild black hole 
$\mathcal{B}:=P_<\times_r S^2$, with $P_<$ denoting the region $r<r_{\rm g}$.
In $\mathcal{B}$, the coordinate vector field $\partial_t$ becomes space-like and $\partial_r$ becomes
time-like. Owing to this, we write for the metric in $\mathcal{B}$ 
\begin{eqnarray}
g
=
-s^{-1}(t){\rm d}t\otimes{\rm d}t+s(t){\rm d}r\otimes{\rm d}r + t^2 w \; .
\end{eqnarray}
Here, $s(t):=|1-r_{\rm g}/t|$ and $w$ denotes the metric on $S^2$, equipped with the usual 
spherical coordinates $(\theta,\phi)$. 

Spatial hypersurfaces $\Sigma$ in $\mathcal{B}$ are conformally flat 
as implied by a vanishing Cotton tensor. 
In order to appreciate conformal flatness, 
it suffices here to consider the region $t\ll r_{\rm g}$ close to the space-like singularity of $\mathcal{B}$,
where the line element of $P_<$ takes the approximate form 
$- (t/r_{\rm g}) {\rm d}t^2 + (r_{\rm g}/t){\rm d}r^2$.  
Following Ehlers and Kundt  \cite{ehl62}, and demanding in addition $\theta\ll 1$, 
the metric can be restated as a type-D Kasner solution characterised by the exponents 
$(p_1,p_2,p_3)=(2/3,2/3,-1/3)$. The corresponding
coordinate transformation is $r  =: (3/2 r_{\rm g})^{1/3} z$, $t =: (9 r_{\rm g}/4)^{1/3} \tau^{2/3}$
and $\theta \exp{({\rm i}\phi)}=: (4/9r_{\rm g})^{1/3} (x+{\rm i} y)$.
In this coordinate neighbourhood, the line element of $\mathcal{B}$ becomes
\begin{equation}
	\label{Kasner}
	{\rm d}s^2 
	=
	-(\mbox{d}\tau)^2 +
	\tau^{4/3}\left((\mbox{d}x)^2+(\mbox{d}y)^2\right)+
	\tau^{-2/3}(\mbox{d}z)^2 \; .
\end{equation}
Harmonic analysis in $\Sigma$ is similar to Euclidean space. 
In particular, the Laplace operator 
in $\Sigma$ factorizes $\Delta_\Sigma = g_{ab}(\tau)\partial^a\partial^b$,
with $\tau$ indexing $\Sigma_\tau$. Generalised eigenfunctions of $\Delta_\Sigma$ are 
plane waves $\exp{({\rm i}g_{_\Sigma}(k,x))}$, where $g_{_\Sigma}$ denotes the induced 
metric tensor in $\Sigma$. 
This coordinate neighbourhood is useful for a quick examination of our results. Moreover,
it allows to relate to the framework suggested by Belinskii, Khalatnikov, and Lifshitz \cite{bel70}.
Let us stress, however, that all the results in this paper have been derived in the usual
Schwarzschild neighbourhood.

\section{Quantum completeness of Schwarzschild black holes}
In this section, we briefly review the argument showing that 
$\mathcal{B}$ is quantum complete. The main result is equation (\ref{oldres}), which has recently been 
published in \cite{sh15}, where a considerably more detailed derivation can be found. Subsequently
it is shown that the free Hamilton density in the ground state 
vanishes towards the classically singular hypersurface $\Sigma_0$.

Since $\mathcal{B}$ is a globally hyperbolic space-time, it is diffeomorphic to $\mathbb{R}\times \Sigma$,
and foliates into spatial hypersurfaces $\Sigma_t$ indexed by Schwarzschild time.
In $\mathcal{B}$, consider the dynamical system $(\mathcal{H},\Phi)$, where $\Phi$ denotes
a real scalar field with Hamilton density $\mathcal{H}=\mathcal{H}_\pi + \mathcal{P}[\Phi]$.
Here, $\mathcal{H}_\pi=\sqrt{-g_{tt}}/2\; \pi^2/{\rm det}(g_{_\Sigma})$ with $\pi=-{\rm i}\delta/\delta\Phi$,
and $\mathcal{P}[\Phi]$ denotes the effective potential.
Quantum completeness refers to free evolution, corresponding to 
$\mathcal{P}[\Phi]=\sqrt{-g_{tt}}/2 g_{_\Sigma}({\rm d}\Phi,{\rm d}\Phi)$, possibly supplemented by a mass term.
The wave functional of the ground state evolves from the initial Cauchy surface $\Sigma_{t_0}$ backwards in time
to $\Sigma_t\; (t\in(0,t_0))$ as $\Psi[\phi](t)=\mathcal{E}(t,t_0)\Psi[\phi](t_0)$, with
\begin{eqnarray}
	\mathcal{E}(t,t_0)
	&=&
	\exp{\left(-{\rm i}\int_{t_0}^t{\rm d}t^\prime
	\int_{{\Sigma_{t^\prime}}}
	{\rm d}\mu_z
	\; \mathcal{H}[\Phi]\right)} 
	\; ,
\end{eqnarray}
where ${\rm d}\mu_z$ denotes the covariant volume form with respect to 
$g_{_{\Sigma}}$, and $z$ refers to the coordinate neighbourhood. 
In $\mathcal{B}$, the evolution operator $\mathcal{E}$ is not unitary. 
Quantum completeness of $\mathcal{B}$ with respect to $(\mathcal{H},\Phi)$
requires $\|\Psi[\phi]\|(t)\le\|\Psi[\phi]\|(t_0) \; \forall t\in(0,t_0)$, implying that
$\mathcal{B}$ can be a sink for probability but not a source, given it is not resolved 
in dynamical degrees of freedom.

We expect that a quadratic functional $\mathcal{K}_2[\phi,\phi](t)$
exists such that 
the wave functional $\Psi^{(0)}[\phi](t)$, corresponding to the ground state of $(\mathcal{H},\Phi)$,
is given by
\begin{eqnarray}
	\Psi^{(0)}[\phi](t)
	&=&
	N^{(0)}(t) \exp{\left(-\mathcal{K}_2[\phi,\phi](t)\right)} \; ,
\end{eqnarray}	
with $N^{(0)}$ denoting the time-dependent normalisation, and $\mathcal{K}_2$ can be expressed in terms 
of the bi-local kernel function $K_2$ as
\begin{eqnarray}
	\mathcal{K}_2[\phi,\phi](t)
	=
	\tfrac{1}{2}
	\int_{{\Sigma_{t}}} {\rm d}\mu_{z_1} {\rm d}\mu_{z_2} \;
	\phi(z_1)K_2(z_1,z_2,t)\phi(z_2) 
	\, .
\end{eqnarray}	
In the vicinity of the Schwarzschild singularity the evolution simplifies considerably, 
\begin{eqnarray}
	\mathcal{E}(t,t_0)
	\rightarrow
	\exp{\left(-\tfrac{{\rm i}c(t_0)}{4M} \ln{t}
	\int_{{\Sigma_{t_0}}}{\rm d}\mu_z\;\tfrac{1}{\sin^2\theta}\tfrac{\delta^2}{\delta\phi^2(z)}
	\right)}\; ,
\end{eqnarray}
where $c(t_0)$ is a constant of integration. As a consequence, the kernel function becomes a contact term in this limit, 
$K_2(z_1,z_2,t)\rightarrow k_2(t)\delta^{(3)}(z_1,z_2)$, which is consistent with the conjecture by 
Belinskii, Khalatnikov and Lifshitz \cite{bel70}: Close to a space-like singularity, the variation 
of observables on $\Sigma_t$ from one location to another becomes irrelevant compared to 
changes in time. Sub-leading corrections to the asymptotic form of $K_2$ deviate from a contact
contribution without changing the qualitative result. 
 
In leading order, the evolution of the wave functional is given by
\begin{eqnarray}
	\label{oldres}
	\lim_{t\rightarrow 0} \Psi^{(0)}[\phi](t)
	=
	\lim_{t\rightarrow 0} \left|\ln\left(t/t_0\right)\right|^{-\Lambda v(\Sigma_t)/2} 
\end{eqnarray}	 
up to constant and phase factors, which are irrelevant for the analysis presented here.
In (\ref{oldres}), an ultra-violet cut-off $\Lambda$ and a volume regularisation $v(\Sigma_t)$
have been introduced. Clearly, the limit $t\rightarrow 0$ is not affected by this simple choice.
Hence, the wave functional has vanishing support towards the Schwarzschild singularity, and
$\|\Psi^{(0)}[\phi]\|(t) \rightarrow 0 \le \|\Psi^{(0)}[\phi]\|(t_0)$ for $t\rightarrow 0$,
as required for $\mathcal{B}$ to be quantum complete with respect to the dynamical system
$(\mathcal{H},\Phi)$. 

Concerning an interpretation:
Consider the set of observables $\mathcal{A}_{\Sigma_t}$ of $(\mathcal{H},\Phi)$ localised on $\Sigma_t$.
Following the logic of geodesic incompleteness, it could be expected that 
an observable $\mathcal{O}_{\Sigma_t}$ exists with an expectation
$\langle \mathcal{O}_{\Sigma_t}\rangle_{\Psi^{(0)}}$ in the ground state that
is ill-defined. 
However, this is not the case since the asymptotic surface $\Sigma_0$ does not support any population
of fields $\phi$, because the associated probability measure vanishes there. 

As an example, consider a free field theory $(\mathcal{H},\Phi)$ in the ground state described by
the Schr\"odinger wave-functional $\Psi^{(0)}[\phi](t)$, where $\phi$ denotes a classical field configuration
over the hypersurface $\Sigma_t$, $t\in(0,2M)$. 
Introducing an auxiliary source $\mathcal{J}$ coupling by 
$\Psi^{(0)}_\mathcal{J}[\phi](t):= \Psi^{(0)}[\phi](t)\exp{(\mathcal{J}[\phi](t))}$ facilitates 
the description of measurement processes.
Observables are evaluated in the ground state $|\Omega\rangle^\mathcal{J}$ 
in the presence of the auxiliary source, which is subsequently set to zero. 
Compositions of the configuration operator are then replaced by the corresponding succession of derivations 
$\delta_\phi$, where
\begin{eqnarray}
	\delta_\phi
	:= 
	\int_{\Sigma_t} {\rm d}\mu_x \; \tfrac{\phi(x)}{\sqrt{{\rm det}(g_{_\Sigma})}}\frac{\delta}{\delta J(x)} \; 
\end{eqnarray}	
is a directional derivative in field space,
with $J$ denoting the ultra-local representation of $\mathcal{J}$. 
For instance, in the presence of an auxiliary source 
\begin{eqnarray}
	&&^{\mathcal{J}}\!\langle\Omega|\Phi^2(\phi)|\Omega\rangle^\mathcal{J} =
	\nonumber \\
	&& 
	\delta_\phi^{\; 2} \exp{\left\{\tfrac{1}{4}\tfrac{1}{\sqrt{{\rm det}(g_{_\Sigma})}}
	\left[{\rm Re}\left(\mathcal{K}_2\right)\right]^{-1}[J,J]
	\right\}}
	\;\mathcal{W}^{(0)}(t) 
\; ,
\end{eqnarray}	
where $\mathcal{W}^{(0)}(t):=\|\Psi^{(0)}[\phi]\|^2(t)$. 
In the absence of the auxiliary source, the ground-state expectation becomes
\begin{eqnarray}
	&&\langle\Omega|\Phi^2(\phi)|\Omega\rangle
	= 
	\nonumber \\
	&&
	\tfrac{1}{2}\sqrt{{\rm det}(g_{_\Sigma})} 
	\left[{\rm Re}\left(\mathcal{K}_2\right)\right]^{-1}[\phi,\phi]
	\; \mathcal{W}^{(0)}(t)
	\; ,
\end{eqnarray}	
which is real and semi-positive definite. In particular, in the vicinity of the limiting 
hypersurface $\Sigma_0$, the ground-state expectation approaches zero, due to the 
temporal support properties associated with the probability density. 
Similarly, it can be 
shown that $\langle\Omega|\pi^2(\phi)|\Omega\rangle$ is real and semi-positive definite, 
and approaches zero towards $\Sigma_0$.
Therefore, the ground-state expectation of $\mathcal{H}$ is in $\mathbb{R}^+$ and vanishes 
towards the would-be singular hypersurface $\Sigma_0$.

\section{Self-interactions}
In this section, polynomial self-interactions are included and their impact on the stability of the ground state
is analysed. For definiteness we consider the effective potential 
$\mathcal{P}_{\rm int}[\Phi]:=\mathcal{P}[\Phi]+ \sqrt{-g_{tt}}\; \lambda\Phi^4/4!$.
The dimensionless coupling $\lambda$ is chosen such that perturbation theory is applicable in 
a neighbourhood of $\Sigma_{t_0}$.
Self-interactions deform the ground-state wave functional away from its Gaussian shape 
\begin{eqnarray}
	\Psi^{(0)}_{\rm int}[\phi](t)
	=
	\Psi^{(0)}[\phi](t)\times \exp{\left(\lambda \mathcal{D}[\phi](t)\right)} \; .
\end{eqnarray}	
The deformation functional $\mathcal{D}=\mathcal{D}_2 + \mathcal{D}_4$
is a sum of the time-dependent nonlinear functionals 
$\mathcal{D}_2: \mathcal{S}^{\otimes 2}\rightarrow \mathcal{C}(\mathbb{R}^+)$ and 
$\mathcal{D}_4: \mathcal{S}^{\otimes 4}\rightarrow \mathcal{C}(\mathbb{R}^+)$, where 
$\mathcal{S}$ denotes the field space and $\mathcal{C}{(\mathbb{R}^+)}$ is the space
of functions depending smoothly on time. As before, local versions can be introduced 
via kernel functions $D_2$ and $D_4$, respectively. Close to the singularity,
$D_j=d_j(t)\Pi_{a=1}^j\delta^{(3)}(z_a)\; , j\in\{2,4\}$, i.e. any spatial information 
close to $\Sigma_0$ is concentrated in a single event. Again, only the temporal gradients matter. 
In this limit, the kernel functions obey the coupled kernel equations
${\rm i}\partial_t \underline{d} = \sqrt{{\rm det}(g)}\, \mathfrak{a}\,\underline{d}$, 
where $\underline{d}:=(d_2,d_4)^{\rm T}$ and $\mathfrak{a}$ is a two-by-two matrix with 
coefficients $\mathfrak{a}_{11}=k(t)$, $\mathfrak{a}_{12}=1$, $\mathfrak{a}_{21}=0$ 
and $\mathfrak{a}_{22}=k(t)$. The asymptotic solution is 
$\underline{d}(t)=(1,1)^{\rm T}/|{\rm ln}(t)|\rightarrow 0$ for $t\rightarrow 0$.

As a consequence, deformations of the Gaussian ground state,
induced by self-interactions, become less and less important towards 
the black-hole singularity, $\mathcal{D}[\phi](t)\rightarrow 0$ for $t\rightarrow 0$. 
In greater detail, asymptotically $\mathcal{D}_j \propto t^{3j/2}/|{\rm ln}(t)|$ for 
$j\in\{2,4\}$ and, hence, 
\begin{eqnarray}
	\lim_{t\rightarrow 0}\Psi^{(0)}_{\rm int}[\phi](t)
	=
	\lim_{t\rightarrow 0}\Psi^{(0)}[\phi](t)
	= 0 \; .
\end{eqnarray}	
%
%
Thus, close to $\Sigma_0$ (i.e. for Schwarzschild times $t\ll t_0$), the dynamical systems
$(\mathcal{H},\Phi)$ and $(\mathcal{H}_\pi,\Phi)$ may be identified. 

This proves that self-interactions cannot cure the classical black-hole singularity 
via back-reaction effects on the external geometry. Close to the singularity 
self-interactions loose their impact on the evolution of the system. The system
becomes asymptotically free, and the stability requirement on the quantum theory
is too stringent to allow the free theory to destabilise even towards $\Sigma_0$. 
Hence, the quantum fields are totally ignorant about the singularity. From this point
of view the classical singularity needs no resolution since it appears as a mathematical 
artefact with no observational consequences whatsoever, assuming the measurements 
are anchored in the framework provided by quantum theory. It seems that 
quantum completeness of the Schwarzschild black-hole protects general relativity 
against its classical incompleteness. In fact, the potential harmful implications 
associated with $\Sigma_0$ decouple from quantum measurements.  

A more abstract reasoning is the following:
Consider classical fields $\phi$ as configurations in $\mathcal{C}^2(\Sigma)$.
In order to ensure a probabilistic interpretation, the Schr\"odinger wave-functionals 
have to be normalisable with respect to some functional measure ${\rm D}\phi$. Wave functionals
enjoying this property can be collected in a state space 
$\mathcal{L}^2(\mathcal{C}^2(\Sigma),{\rm D}\phi)$, which obviously requires a mathematical
justification beyond the scope of this article. Even for these wave functionals, $\mathcal{H}_\pi$
is not self-adjoint, but the spectrum contains only functions with a positive-semidefinite imaginary 
part. As a consequence, towards the singularity $\mathcal{E}(t,t_0)$ becomes exponentially damped.
Self-interactions cannot harm this regularisation of the classical singularity, simply because 
they are given as compositions of multiplication operators. Furthermore, $\mathcal{H}\rightarrow\mathcal{H}_\pi$
towards the singularity, where the limit is taken in a generalisation of the strong operator topology
appropriate for the functional calculus involved here. From this point of view, a self-interacting 
quantum probe is totally ignorant about the classical singularity.

\section{Excitations}
Excitations of the ground state are not an integral component in the definition of
quantum completeness. If the dynamical system $(\mathcal{H},\Phi)$ is unstable, then
excitations might trigger a transition towards a stable ground state. 

The ground state is an eigenstate of the conjugated momentum field,
$(\pi(t,x) - {\rm i}\delta\mathcal{K}_2/\delta\phi(t,x))\Psi^{(0)}[\phi](t)=0$,
and a kernel element of the operator valued functional $a[f](t)$, describing the 
absorption of a field $f\in\mathcal{S}_{\rm os}$, where the index 'os' implies 
the restriction to on-shell fields.  
The above eigenstate equation is a ultra-local version of
absorption. Emission can be considered accordingly using the adjoint $a^\dagger[f](t)$.
As usual, on $\Sigma_t$ the following algebraic relation holds: 
\begin{eqnarray}
	\label{alg}
	\left[a[f](t),a^\dagger[f^\prime](t)\right]
	=
	2{\rm Re}\left(\mathcal{K}_2[f,f^\prime]\right)(t)
	\; . 
\end{eqnarray}	
An excitation relative to the ground state is given by
$\Psi^{(1)}[f,\phi](t) := a^\dagger[f](t) \Psi^{(0)}[\phi](t)$. Note that $\phi\in\mathcal{S}$,
while $f\in\mathcal{S}_{\rm os}\subset\mathcal{S}$, i.e. the emission operator creates
on-shell information and stores it in the excited state
$\Psi^{(1)}[f,\phi](t)=2{\rm Re}\left(\mathcal{K}_2[f,\phi]\right)(t)\Psi^{(0)}[\phi](t)$. 
Therefore, exciting the ground state by emitting an on-shell quantum simply results in 
a functional renormalisation of the ground state. Owing to the algebraic relation (\ref{alg}),
we find 
\begin{eqnarray}
	\left\|\Psi^{(1)}\right\|^2(t)
	=
	2{\rm Re}\left(\mathcal{K}_2[f,f]\right)(t)
	\left\|\Psi^{(0)}\right\|^2(t)
	\; .
\end{eqnarray}
So quantum completeness of the ground state
is a necessary but not sufficient criterion for the stability of the first excited state. 
In addition, ${\rm Re}(\mathcal{K}_2[f,f])(t)<\infty$ is required for all $f\in\mathcal{S}_{\rm os}$.

For vanishing Schwarzschild time, the renormalisation becomes constant and is therefore inconsequential. 
This can be seen as follows: Up to sub-leading contributions, the time dependence of $f(t,x)=T(t)R(x)$
is given by $(t\partial_t^2 + \partial_t)T=\kappa T$ with $\kappa$ a constant determined by the equation for $R$.
For vanishing Schwarzschild time, $T$ should be singular. Introducing $\tau:=\zeta t$, and taking the limit
$t\rightarrow 0$, $\zeta\rightarrow\infty$ such that the rescaled Schwarzschild time $\tau$ remains constant,
the equation of motion for $T$ becomes $(\tau\partial_\tau^2 + \partial_\tau)T=0$. Thus, up to an additive
constant, $T={\rm ln}(t)$. Therefore, 
${\rm Re}(\mathcal{K}_2[f,f^\prime])=$ const, because the time dependence of the corresponding kernel 
function cancels exactly against the time dependence of the mode functions and the volume form.
Note that the additive constant poses no problem due to the prescription for taking the asymptotic limit.
As a consequence, $\|\Psi^{(1)}\|^2\rightarrow 0$ towards the black-hole singularity.  
In fact, as can be seen by induction, all excitations $\Psi^{(n)}\, (n\in\mathbb{N})$ give rise to 
a vanishing probability measure on $\Sigma_0$. Neither the ground state nor any excited states are populated
with fields on $\Sigma_0$. The natural probability measure protects the stability of any state, and this 
stability protection can be traced back to a persistent ground state.

\section{Discussion}
In this article the notion of quantum mechanical completeness is adapted to situations 
where the only adequate description is in terms of (interacting) quantum fields in 
dynamical space-times. 
The adaption necessarily generalises from requiring a
unitary evolution by demanding a normalisation condition that ensures a probabilistic 
interpretation. Of course, this condition reduces to unitarity in the absence of
dynamical sources. 
While originally stated for free fields in a Gaussian ground-state,
it is shown to extend to interacting quantum fields in arbitrary states. 
It is tempting to expect that this extension is rather trivial if the ground state
admits a Gaussian wave functional. This expectation has to be confronted with the dynamics 
of the external space-time that sources different terms in the Hamiltonian differently.
%
It is important to stress that both, geodesic and quantum completeness, assume a 
background space-time, which is either diagnosed by test particles or by test fields, respectively.
This assumption, however, can only be investigated in a quantum theory of fields.

Whether a given dynamics is consistent with a probabilistic interpretation is usually 
examined in an asymptotic framework pertinent to scattering theory. There the stability 
of the ground state is studied in the presence of an external source
after an infinite amount of time has passed. This is clearly not an option in arbitrary 
space-times. Furthermore, it seems intuitive that stability challenges 
are anchored in the vicinity of space-time singularities, which suggests 
a more local stability analyses. For these reasons, the Schr\"odinger representation 
of quantum field theory is quite convenient, which allows, in particular, to investigate the
stability of a given quantum system in a dynamical space-time after a finite amount of time elapsed.

The Schr\"odinger representation requires a functional generalisation of many quantum-mechanical 
concepts. In particular, choosing the configuration field as the multiplication operator, 
the associated momentum field becomes a functional derivative. And the norm of a wave functional 
requires a functional integral over the configuration fields. Many of these functional techniques 
can be disputed on mathematical grounds. However, the stability analysis is entirely at the
qualitative level and not based on any specific regularisation. 

%
The main result of this article is that Schwarzschild black-holes are quantum complete, which has
a very precise meaning. However, equally precise they are qualified as geodesically 
incomplete space-times
by the singularity theorem of Hawking and Penrose. 
Of course, both completeness notions are logically consistent within their respective domains of validity.
If we are to derive further consequences from these notions, in particular concerning the consistency
of black holes and of general relativity, it is important to understand which domain and therefore
which completeness notion is applicable given the physical conditions.
Our point of view advocated here is the following:
Geodesic completeness is a concept in the category of smooth manifolds as models for space-times.
To the extent that we can be certain that these models can be probed by physical events it is falsifiable. 
In the vicinity of space-like singularities, spatial correlations become trivial, i.e.~events can only
be spatially correlated if they are stacked on top of each other. As might be expected, what matters 
in the vicinity of a space-like singularity are temporal correlations. In fact, temporal gradients
correspond to a characteristic length scale that is smaller than the length scale characterising 
the spatial extent of any conceivable classical measurement device. Therefore, any completeness 
diagnosis based on classical measurements is inappropriate given the physical conditions. 
Any measurement process in the vicinity 
of a black-hole singularity has to rely on quantum field theory. 
In the context of classical singularity 
theorems, the only falsifiable completeness notion applicable to black-hole interiors is 
quantum completeness.

This argument is not in conflict with the logic underlying the usual quantisation prescription,
precisely because the probability measure is always well-defined. In particular, the Gaussian 
ground-state is respected by self-interactions, provided the system was in a weak-coupling regime
initially. This is in accordance with the intuitive expectation that the free dynamics (temporal correlations)
dominates in the vicinity of the singularity. Consequently, excitations relative to the ground state 
cannot change the conclusion. 
Let us stress that these results are in full accordance with the dynamical stability of classical
field configuration in Schwarzschild space-time, as has been established in \cite{jen86,eli87,eli88}. 
Temporal support for field configurations is strictly restricted to the interval $(0,t_0]$ with the 
initial time $t_0<r_{\rm g}$, and the field configurations are smooth on this interval.
 
%
Black-hole interiors are quantum complete and this notion is sensible from a physics point of view
even in the vicinity of the classically singular hypersurface. In contrast, geodesic incompleteness of black holes,
albeit a mathematical rigorous qualification, is not a physical statement since any operative measurement 
has to employ physics beyond point particle dynamics. As a consequence, the classically singular hypersurface
bears no impact on observables based on bookkeeping devices (fields) with sensible dynamics.
Less sensible is the argument
that geometrical observables such as the Kretschmann scalar would diverge at the origin. This line of 
argument is already invalidated for simple bound-state problems in quantum mechanics, for instance 
the hydrogen atom. Clearly, the Coulomb potential enjoys geodesic and potential incompleteness,
which is inconsequential for hydrogen as a quantum bound-state. Albeit the singular structure in this case
is just a point, quantum completeness is established by arguments related to the support properties 
of the probability measure, as well. In the case of black holes, the singular structure is space-like,
but corresponds to a limiting instant in time. 

\begin{acknowledgments}
It is a great pleasure to thank Ingemar Bengtsson, Kristina Giesel, Maximilian K\"ogler, Florian Niedermann, 
and Maximilian Urban for delightful discussions. We thank Robert C. Myers for sharing his ideas about this topic. 
We appreciate financial support of our work
by the DFG cluster of excellence 'Origin and Structure of the Universe', the Humboldt Foundation,
and by TRR 33 'The Dark Universe'. 

\end{acknowledgments}



\bibliography{literww}

\begin{thebibliography}{12}%
\makeatletter
\providecommand \@ifxundefined [1]{%
 \@ifx{#1\undefined}
}%
\providecommand \@ifnum [1]{%
 \ifnum #1\expandafter \@firstoftwo
 \else \expandafter \@secondoftwo
 \fi
}%
\providecommand \@ifx [1]{%
 \ifx #1\expandafter \@firstoftwo
 \else \expandafter \@secondoftwo
 \fi
}%
\providecommand \natexlab [1]{#1}%
\providecommand \enquote  [1]{``#1''}%
\providecommand \bibnamefont  [1]{#1}%
\providecommand \bibfnamefont [1]{#1}%
\providecommand \citenamefont [1]{#1}%
\providecommand \href@noop [0]{\@secondoftwo}%
\providecommand \href [0]{\begingroup \@sanitize@url \@href}%
\providecommand \@href[1]{\@@startlink{#1}\@@href}%
\providecommand \@@href[1]{\endgroup#1\@@endlink}%
\providecommand \@sanitize@url [0]{\catcode `\\12\catcode `\$12\catcode
  `\&12\catcode `\#12\catcode `\^12\catcode `\_12\catcode `\%12\relax}%
\providecommand \@@startlink[1]{}%
\providecommand \@@endlink[0]{}%
\providecommand \url  [0]{\begingroup\@sanitize@url \@url }%
\providecommand \@url [1]{\endgroup\@href {#1}{\urlprefix }}%
\providecommand \urlprefix  [0]{URL }%
\providecommand \Eprint [0]{\href }%
\providecommand \doibase [0]{http://dx.doi.org/}%
\providecommand \selectlanguage [0]{\@gobble}%
\providecommand \bibinfo  [0]{\@secondoftwo}%
\providecommand \bibfield  [0]{\@secondoftwo}%
\providecommand \translation [1]{[#1]}%
\providecommand \BibitemOpen [0]{}%
\providecommand \bibitemStop [0]{}%
\providecommand \bibitemNoStop [0]{.\EOS\space}%
\providecommand \EOS [0]{\spacefactor3000\relax}%
\providecommand \BibitemShut  [1]{\csname bibitem#1\endcsname}%
\let\auto@bib@innerbib\@empty
\bibitem [{\citenamefont {Hawking}\ and\ \citenamefont
  {Penrose}(1970)}]{hawk70}%
  \BibitemOpen
  \bibfield  {author} {\bibinfo {author} {\bibfnamefont {S.~W.}\ \bibnamefont
  {Hawking}}\ and\ \bibinfo {author} {\bibfnamefont {R.}~\bibnamefont
  {Penrose}},\ }\href@noop {} {\bibfield  {journal} {\bibinfo  {journal}
  {Proceedings of the Royal Society of London A}\ }\textbf {\bibinfo {volume}
  {314}},\ \bibinfo {pages} {529} (\bibinfo {year} {1970})}\BibitemShut
  {NoStop}%
\bibitem [{\citenamefont {Reed}\ and\ \citenamefont {Simon}(1975)}]{reed75}%
  \BibitemOpen
  \bibfield  {author} {\bibinfo {author} {\bibfnamefont {M.}~\bibnamefont
  {Reed}}\ and\ \bibinfo {author} {\bibfnamefont {B.}~\bibnamefont {Simon}},\
  }\href@noop {} {\emph {\bibinfo {title} {Methods of modern mathematical
  physics II: Fourier Analysis, Self-Adjointness}}},\ Vol.~\bibinfo {volume}
  {2}\ (\bibinfo  {publisher} {Elsevier},\ \bibinfo {year} {1975})\BibitemShut
  {NoStop}%
\bibitem [{\citenamefont {Horowitz}\ and\ \citenamefont
  {Marolf}(1995)}]{hor95}%
  \BibitemOpen
  \bibfield  {author} {\bibinfo {author} {\bibfnamefont {G.}~\bibnamefont
  {Horowitz}}\ and\ \bibinfo {author} {\bibfnamefont {D.}~\bibnamefont
  {Marolf}},\ }\href@noop {} {\bibfield  {journal} {\bibinfo  {journal}
  {Physical Review D}\ }\textbf {\bibinfo {volume} {52}},\ \bibinfo {pages}
  {5670} (\bibinfo {year} {1995})}\BibitemShut {NoStop}%
\bibitem [{\citenamefont {Ashtekar}\ and\ \citenamefont
  {Magnon}(1975)}]{ash75}%
  \BibitemOpen
  \bibfield  {author} {\bibinfo {author} {\bibfnamefont {A.}~\bibnamefont
  {Ashtekar}}\ and\ \bibinfo {author} {\bibfnamefont {A.}~\bibnamefont
  {Magnon}},\ }in\ \href@noop {} {\emph {\bibinfo {booktitle} {Proceedings of
  the Royal Society of London A: Mathematical, Physical and Engineering
  Sciences}}},\ Vol.\ \bibinfo {volume} {346}\ (\bibinfo {organization} {The
  Royal Society},\ \bibinfo {year} {1975})\ pp.\ \bibinfo {pages}
  {375--394}\BibitemShut {NoStop}%
\bibitem [{\citenamefont {Wald}(1978)}]{wald78}%
  \BibitemOpen
  \bibfield  {author} {\bibinfo {author} {\bibfnamefont {R.~M.}\ \bibnamefont
  {Wald}},\ }\href@noop {} {\bibfield  {journal} {\bibinfo  {journal} {Journal
  of Mathematical Physics}\ }\textbf {\bibinfo {volume} {20}} (\bibinfo {year}
  {1978})}\BibitemShut {NoStop}%
\bibitem [{Note1()}]{Note1}%
  \BibitemOpen
  \bibinfo {note} {This corresponds to relaxing the requirements posed on
  evolution generators to linear unbounded operators that are closed, have a
  dense domain, whose resolvent set contains all positive reals and whose
  resolvent operator, which can be thought of as the Laplace transform of the
  semigroup, is bounded from above for each element of the resolvent set by its
  inverse.}\BibitemShut {Stop}%
\bibitem [{\citenamefont {Hofmann}\ and\ \citenamefont
  {Schneider}(2015)}]{sh15}%
  \BibitemOpen
  \bibfield  {author} {\bibinfo {author} {\bibfnamefont {S.}~\bibnamefont
  {Hofmann}}\ and\ \bibinfo {author} {\bibfnamefont {M.}~\bibnamefont
  {Schneider}},\ }\href {\doibase 10.1103/PhysRevD.91.125028} {\bibfield
  {journal} {\bibinfo  {journal} {Phys. Rev. D}\ }\textbf {\bibinfo {volume}
  {91}},\ \bibinfo {pages} {125028} (\bibinfo {year} {2015})}\BibitemShut
  {NoStop}%
\bibitem [{\citenamefont {Ehlers}\ and\ \citenamefont {Kundt}(1962)}]{ehl62}%
  \BibitemOpen
  \bibfield  {author} {\bibinfo {author} {\bibfnamefont {J.}~\bibnamefont
  {Ehlers}}\ and\ \bibinfo {author} {\bibfnamefont {W.}~\bibnamefont {Kundt}},\
  }in\ \href@noop {} {\emph {\bibinfo {booktitle} {The Theory of
  Gravitation}}}\ (\bibinfo  {publisher} {John Wiley \& Sons, Inc.},\ \bibinfo
  {year} {1962})\ pp.\ \bibinfo {pages} {49--101}\BibitemShut {NoStop}%
\bibitem [{\citenamefont {Belinskii}\ \emph {et~al.}(1970)\citenamefont
  {Belinskii}, \citenamefont {Khalatnikov},\ and\ \citenamefont
  {Lifshitz}}]{bel70}%
  \BibitemOpen
  \bibfield  {author} {\bibinfo {author} {\bibfnamefont {V.}~\bibnamefont
  {Belinskii}}, \bibinfo {author} {\bibfnamefont {I.}~\bibnamefont
  {Khalatnikov}}, \ and\ \bibinfo {author} {\bibfnamefont {E.}~\bibnamefont
  {Lifshitz}},\ }\href {\doibase 10.1080/00018737000101171} {\bibfield
  {journal} {\bibinfo  {journal} {Advances in Physics}\ }\textbf {\bibinfo
  {volume} {19}},\ \bibinfo {pages} {525} (\bibinfo {year} {1970})}\BibitemShut
  {NoStop}%
\bibitem [{\citenamefont {Jensen}\ and\ \citenamefont
  {Candelas}(1986)}]{jen86}%
  \BibitemOpen
  \bibfield  {author} {\bibinfo {author} {\bibfnamefont {B.~P.}\ \bibnamefont
  {Jensen}}\ and\ \bibinfo {author} {\bibfnamefont {P.}~\bibnamefont
  {Candelas}},\ }\href {\doibase 10.1103/PhysRevD.33.1590} {\bibfield
  {journal} {\bibinfo  {journal} {Phys. Rev. D}\ }\textbf {\bibinfo {volume}
  {33}},\ \bibinfo {pages} {1590} (\bibinfo {year} {1986})}\BibitemShut
  {NoStop}%
\bibitem [{\citenamefont {Elizalde}(1987)}]{eli87}%
  \BibitemOpen
  \bibfield  {author} {\bibinfo {author} {\bibfnamefont {E.}~\bibnamefont
  {Elizalde}},\ }\href {\doibase 10.1103/PhysRevD.36.1269} {\bibfield
  {journal} {\bibinfo  {journal} {Phys. Rev. D}\ }\textbf {\bibinfo {volume}
  {36}},\ \bibinfo {pages} {1269} (\bibinfo {year} {1987})}\BibitemShut
  {NoStop}%
\bibitem [{\citenamefont {Elizalde}(1988)}]{eli88}%
  \BibitemOpen
  \bibfield  {author} {\bibinfo {author} {\bibfnamefont {E.}~\bibnamefont
  {Elizalde}},\ }\href {\doibase 10.1103/PhysRevD.37.2127} {\bibfield
  {journal} {\bibinfo  {journal} {Phys. Rev. D}\ }\textbf {\bibinfo {volume}
  {37}},\ \bibinfo {pages} {2127} (\bibinfo {year} {1988})}\BibitemShut
  {NoStop}%
\end{thebibliography}%

\end{document}